\documentclass[aps,prd,preprint,tightenlines,superscriptaddress,
   showpacs,nofootinbib,12pt]{revtex4-1}
\usepackage{amsmath}
\usepackage{graphicx}

\begin{document}

\title{Constraints from neutrino masses and muon (g-2) in the
bilinear R-parity violating supersymmetric model}

\author{R. S. Hundi}
\email{tprsh@iacs.res.in}
\affiliation{
Department of Theoretical Physics,
Indian Association for the Cultivation of Science,
2A $\&$ 2B Raja S.C. Mullick Road,
Kolkata - 700 032, India.}

\begin{abstract}
Bilinear R-parity violating supersymmetric model is a viable model
which can explain the smallness of neutrino masses
and the mixing pattern in the lepton sector. In this model, there
is a common set of parameters which determine the neutralino
and chargino masses, also determine the neutrino masses
and the anomalous magnetic moment of muon, $(g-2)_\mu$. From the experimental data
on neutrino masses and mixing angles, and also from the
fact that the experimentally measured value of $(g-2)_\mu$ differs from
the corresponding standard model value by 3$\sigma$, we have
analyzed some constraints on these model parameters.
Constraints on these model parameters are obtained for some
values of supersymmetry breaking soft parameters.
These constraints can set upper and lower bounds on
the chargino masses of this model.
\end{abstract}

\maketitle

\section{Introduction}
\label{s:intro}

Since the
discovery of neutrinos, their mass estimation has provoked much excitement
among particle phenomenologists. The deficit in the neutrino flux
from solar \cite{solar} and atmospheric \cite{atmos} neutrinos
have confirmed that at least two neutrinos
should have non-zero masses. The recent global fitting to the
various neutrino experimental data has given the following
mass-squared differences and mixing angle values at 1$\sigma$ level \cite{STV}.
\begin{eqnarray}
\Delta m_{21}^2 = (7.65_{-0.20}^{+0.23})\times 10^{-5}~{\rm eV}^2,
\quad
|\Delta m_{31}^2| = (2.40_{-0.11}^{+0.12})\times 10^{-3}~{\rm eV}^2,
\nonumber \\
\sin^2\theta_{12}=0.304_{-0.016}^{+0.022},\quad
\sin^2\theta_{23}=0.50_{-0.06}^{+0.07},\quad
\sin^2\theta_{13}=0.01_{-0.011}^{+0.016},
\label{eq:numnm}
\end{eqnarray}
where $\Delta m_{ij}^2=m_i^2-m_j^2$, $m_i$ being the mass eigenvalue
of a neutrino field for $i=1,2,3$. The various $\theta$s
given above are the mixing angles in the neutrino sector.
Apart from the above data, cosmological
observations would give an upper bound on the neutrino mass scale
to be of the order of 0.1 eV \cite{cosmos}. Such a low mass among the known
particle masses can be explained through models based on seesaw
mechanism \cite{seesaw}.
A consequence of seesaw mechanism is the existence of neutrinoless double
beta decay process \cite{2beta}, which is being investigated in experiments.
The mixing angles in eq. (\ref{eq:numnm})
suggest that
there should be two large and one small angles
in the neutrino sector, whose pattern is very much different
from that in the quark sector where all the three
mixing angles are small. This issue also indicates new
physics for which a model has been proposed
\cite{AEK}. Moreover, the central values
of the three mixing angles in eq. (\ref{eq:numnm}) are
very close to the tri-bimaximal pattern which is given
below \cite{HPS}
\begin{equation}
\sin\theta^{\rm TB}_{12}=\frac{1}{\sqrt{3}},\quad
\sin\theta^{\rm TB}_{23}=\frac{1}{\sqrt{2}},\quad
\sin\theta^{\rm TB}_{13}=0.
\label{eq:tbm}
\end{equation}
Several models based on discreet symmetries have been motivated
to explain the tri-bimaximal pattern in the neutrino sector \cite{A4}.

On the other hand, the anomalous magnetic moment of muon
is one of the most precisely measured quantities in
experiments, which can be quantified
as $a_\mu=(g-2)_\mu/2$. After the recent experiment E821 at the Brookhaven
National Laboratory, a world average of \cite{g-2exp}
\begin{equation}
a_\mu^{\rm EXP} = 11659208.0(6.3)\times 10^{-10}
\end{equation}
has been achieved with a precision of 0.54 parts per million.
Various groups have computed the theoretical value of $a_\mu$
in the standard model (SM), i.e. $a^{\rm SM}_\mu$. The uncertainty
in computing the $a^{\rm SM}_\mu$ is largely coming from
the hadronic loop corrections.
For a review on the calculational methods and results of
$a^{\rm SM}_\mu$, see \cite{g-2rev,re-evl}. The results from these
groups indicate that the experimental value $a^{\rm EXP}_\mu$
differs from the theoretically calculated value $a^{\rm SM}_\mu$
by about $2-3~\sigma$. Although some work needs
to be done to reduce the size of error bars on the
$a_\mu$ in the SM, at this moment it is
tempting to say that this difference
indicates physics beyond the SM.
In this work, we have taken the difference between the experiment
and the SM value of $a_\mu$ as \cite{DE}
\begin{equation}
\Delta a_\mu = a_\mu^{\rm EXP} - a_\mu^{\rm SM} =
(27.7\pm 9.3)\times 10^{-10}.
\label{eq:damu}
\end{equation}

Supersymmetry \cite{susy} is a leading candidate for physics beyond the
SM. In some class of supersymmetric models where
the lepton number is assumed to be violated,
the neutrino masses and mixing angles can be understood.
Among these, bilinear R-parity violating (BRpV) supersymmetric
model \cite{brpvrev} has rich phenomenology \cite{brpv-phe}
and simple structure to explain the neutrino masses and mixing
angles \cite{DHPRV,DL,GR}. In the
BRpV supersymmetric model, a single neutrino acquires non-zero mass at the
tree level through a mixing between flavor neutrinos
and neutralinos. The remaining two neutrinos acquire masses
at 1-loop level, among which the dominant diagrams
are generated through the neutralino$-$sneutrino$-$Higgs
loops. Since neutralinos are playing a part in giving masses
to the neutrinos, the
parameters of neutralinos obviously determine the neutrino masses
in this model. The neutrino masses generated at 1-loop level also depend
on some supersymmetry breaking soft parameters due to the presence
of sneutrinos and Higgses
in the loop. Interestingly, the
parameters related to neutralinos and sneutrino masses also
determine the $(g-2)_\mu$ in the BRpV supersymmetric model. This is due to the
fact that the leading supersymmetric contribution to the
$(g-2)_\mu$ in this model comes from 1-loop diagrams generated by
neutralino$-$smuon and chargino$-$muon-sneutrino loops, where the
chargino masses are determined by the parameters of neutralino
matrix. Since both
the neutrino mass-squares and the $(g-2)_\mu$ have been measured
precisely, it is interesting to see if these two set of
observable quantities can put
constraints on the above said parameters in the BRpV supersymmetric
model, which is the aim of this work.

The paper has been organized as follows. In Sec. \ref{sec:feau}
we briefly describe the BRpV supersymmetric model and how neutrinos acquire
masses at tree and 1-loop levels. We also briefly explain the
supersymmetric contribution from the BRpV model to the
$(g-2)_\mu$. In Sec. \ref{sec:AnR} we describe our method of
analyzing constraints on the model parameters due to the
neutrino masses and the $(g-2)_\mu$, and also present
our results. We conclude in Sec. \ref{sec:con}.

\section{The model, neutrino masses and $(g-2)_\mu$}
\label{sec:feau}

The superpotential of the BRpV supersymmetric
model is
\begin{equation}
W=Y_u^{ij}\hat{Q}_i\hat{U}_j\hat{H}_u - Y_d^{ij}\hat{Q}_i\hat{D}_j\hat{H}_d
- Y_e^{ij}\hat{L}_i\hat{E}_j\hat{H}_d + \mu\hat{H}_u\hat{H}_d
+ \epsilon_i\hat{L}_i\hat{H}_u,
\label{eq:supbi}
\end{equation}
where the indices $i,j$ run from 1 to 3. The superfields
$\hat{Q}$, $\hat{U}$ and $\hat{D}$ are doublet,
singlet up-type and singlet down-type quark fields,
respectively. $\hat{L}$ and $\hat{E}$ are doublet
and singlet charged lepton superfields, respectively. $\hat{H}_u$
and $\hat{H}_d$ are up- and down-type Higgs superfields,
respectively.
The bilinear term $\hat{H}_u\hat{H}_d$
is called $\mu$-term and the mass parameter
$\mu$ is assumed to be ${\cal O}$(100) GeV, in order to have theoretical
consistency of the model.
The other bilinear term $\hat{L}\hat{H}_u$
violates lepton number as well as R-parity and it is called BRpV term.
Along with the supersymmetry conserving terms of eq. (\ref{eq:supbi}),
we also get soft terms
in the low energy regime as a result of supersymmetry breaking
in the high energy scale.
Below we give those soft terms in the scalar potential, which are necessary for both
the neutrino masses and the $(g-2)_\mu$.
\begin{eqnarray}
V_{\rm soft}&=&\frac{1}{2}M_1\tilde{B}\tilde{B}+\frac{1}{2}M_2\tilde{W}\tilde{W}+
(m_L^2)_{ij}\tilde{L}^*_i\tilde{L}_j + (m_E^2)_{ij}\tilde{E}^*_i\tilde{E}_j
+ (A_EY_E)_{ij}\tilde{L}_i\tilde{E}_jH_d
\nonumber \\
&&+(b_\mu)H_uH_d +(b_\epsilon)_i\tilde{L}_iH_u + \cdots,
\label{eq:soft}
\end{eqnarray}
where the first two terms in the above equation are mass terms
for U(1)$_Y$ and SU(2)$_L$ gaugino fields, respectively.
The third and fourth terms in the above equation are soft scalar
mass terms for the left-handed slepton doublet and right-handed
slepton singlet fields, respectively, and the fifth term is called
$A$-term for slepton fields. The
last two terms in the above equation arises due to the $\mu$-term
and the BRpV term of eq. (\ref{eq:supbi}), respectively.

The BRpV term of eq. (\ref{eq:supbi}) allows mixing mass terms between
the up-type neutral higgsino field ($\tilde{H}_u^0$) and the three left-handed
neutrino fields ($\nu_i$). Also, since lepton number is violated
by the BRpV term, the scalar components of left-handed neutrino
superfields acquire non-zero vacuum expectation values (vevs).
To simplify our formulas, we work in a basis where the vevs of
sneutrino fields are zero. For $\psi_N = (\tilde{B},
\tilde{W}^3,\tilde{H}_u^0,\tilde{H}_d^0,\nu_1,\nu_2,\nu_3)^T$, at the
tree level we get the following mixing masses:
${\cal L} = -\frac{1}{2}\psi^T_N M_N \psi_N + {\rm h.c.}$,
where
\begin{equation}
M_N=
\left(\begin{array}{cc}
M_{\chi^0} & m \\
m^T & 0
\end{array}\right),
\label{eq:n0}
\end{equation}
\begin{equation}
M_{\chi^0}=\left(\begin{array}{cccc}
M_1 & 0 & \frac{1}{\sqrt{2}}g_1v_u & -\frac{1}{\sqrt{2}}g_1v_d \\
0 & M_2 & -\frac{1}{\sqrt{2}}g_2v_u & \frac{1}{\sqrt{2}}g_2v_d \\
\frac{1}{\sqrt{2}}g_1v_u & -\frac{1}{\sqrt{2}}g_2v_u & 0 & -\mu \\
-\frac{1}{\sqrt{2}}g_1v_d & \frac{1}{\sqrt{2}}g_2v_d & -\mu & 0
\end{array}\right),
\quad
m = \left(\begin{array}{ccc}
0 & 0 & 0 \\
0 & 0 & 0 \\
\epsilon_1 & \epsilon_2 & \epsilon_3 \\
0 & 0 & 0
\end{array}\right).
\label{eq:mchim}
\end{equation}
Here, $g_1,g_2$ are the gauge couplings corresponding to U(1)$_Y$ and SU(2)$_L$,
respectively.
For the vevs of Higgs scalar fields we have followed the convention:
$\langle H_d^0 \rangle = v_d = v\cos\beta,\langle H_u^0 \rangle = v_u
= v\sin\beta$,
where $v=174$ GeV is the electroweak scale. For $\epsilon_i <<
m_{\rm susy}\sim$ TeV, the left-handed neutrinos acquire non-zero masses
through a seesaw mechanism at around TeV scale. At the tree level,
upto leading order in the expansion of $\frac{1}{m_{\rm susy}}$,
the neutrino masses would be determined by
$m_\nu = -m^T M^{-1}_{\chi^0} m$.
It can be easily seen that after plugging eq. (\ref{eq:mchim})
in this expression, only one neutrino mass eigenvalue would be non-zero.
The other two neutrino eigenstates get masses at 1-loop level, for
which many different possible loop diagrams exist.
However, it has been argued
in Ref. \cite{GR} that in the case where the tree level neutrino
mass eigenvalue is dominant, the
1-loop generated by two insertions of $b_\epsilon$ would
be the dominant one among all the loop diagrams for the neutrinos,
which is the case we consider in this work.

For simplicity, we assume that
the sneutrinos are degenerate. In such a case the dominant
contribution to the neutrino mass matrix can be written as \cite{DL,GR}
\begin{equation}
(m_\nu)_{ij}=a_0\epsilon_i\epsilon_j+a_1(b_\epsilon)_i(b_\epsilon)_j,
\label{eq:tn1l}
\end{equation}
where the indices $i,j$ run from 1 to 3. The first term in the
above equation is due to the tree level effect and the second term
is from the 1-loop diagram. The expressions for $a_0$ and $a_1$ are \cite{DL,GR}
\begin{eqnarray}
a_0&=&\frac{m_Z^2m_{\tilde{\gamma}}\cos^2\beta}{\mu(m_Z^2m_{\tilde{\gamma}}
\sin 2\beta-M_1M_2\mu)},\quad m_{\tilde{\gamma}}=\cos^2\theta_W M_1+\sin^2\theta_W M_2,
\nonumber \\
a_1&=&\sum_{i = 1}^{4}\frac{(g_2(U_0)_{2i}-g_1(U_0)_{1i})^2}
{4\cos^2\beta}(m_{N^0})_i\left(I_4(m_h,m_{\tilde \nu},m_{\tilde \nu},(m_{N^0})_i)\cos^2(\alpha-\beta)
\right.
\nonumber \\
&& \left. +I_4(m_H,m_{\tilde \nu},m_{\tilde \nu},(m_{N^0})_i)\sin^2(\alpha-\beta)
-I_4(m_A,m_{\tilde \nu},m_{\tilde \nu},(m_{N^0})_i)\right),
\label{eq:a0a1}
\end{eqnarray}
where $m_Z$ is the $Z$ boson mass, $\theta_W$ is the Weinberg angle,
and the $m_h$, $m_H$ and $m_A$ are the light, heavy and
pseudo-scalar Higgs boson masses, respectively. The unitary matrix $U_0$
diagonalizes the neutralino matrix $M_{\chi^0}$ of
eq. (\ref{eq:mchim}) and the $(m_{N^0})_i$ are the neutralino
mass eigenvalues. $m_{\tilde \nu}$ is the mass of sneutrino
field. The angle $\alpha$ is the mixing angle
in the rotation matrix which diagonalizes the light and
heavy Higgs boson fields. It is to be noted that the Higgs boson
masses depend on the soft parameter $b_\mu$ which
is given in eq. (\ref{eq:soft}).
The function $I_4$ is given by
\begin{eqnarray}
I_4(m_1,m_2,m_3,m_4)&=&\frac{1}{m_1^2-m_2^2}[I_3(m_1,m_3,m_4)-
I_3(m_2,m_3,m_4)],
\nonumber \\
I_3(m_1,m_2,m_3)&=&\frac{1}{m_1^2-m_2^2}[I_2(m_1,m_3)-I_2(m_2,m_3)],
\nonumber \\
I_2(m_1,m_2)&=&-\frac{1}{16\pi^2}\frac{m_1^2}{m_1^2-m_2^2}\ln\frac{m_1^2}{m_2^2}.
\label{eq:i4}
\end{eqnarray}
Many of the mass parameters appearing in the above neutrino mass
expression are of order a few 100 GeV,
except for $\epsilon_i$ and $(b_\epsilon)_i$. Taking all the
supersymmetric mass scales to be ${\cal O}$(100 GeV) and $\tan\beta
\sim {\cal O}(10)$, the scales of $\epsilon$ and $b_\epsilon$
should be ${\cal O}(10^{-3})$ GeV and ${\cal O}(0.1)$ GeV$^2$,
respectively, in order to get a tiny neutrino mass scale of 0.1 eV.
In this order of estimation we have taken
into account of the partial cancellation of the Higgs boson
contributions in the 1-loop \cite{GR}.
The smallness of the estimated parameters, $\epsilon$ and $b_\epsilon$, can be
motivated in a supergravity setup \cite{HPT} and also through
other mechanisms \cite{SRp}.

Now, we explain the supersymmetric contribution to the $a_\mu$ in the BRpV
supersymmetric model, which we quantify as $\Delta a_\mu$. The forms of the
leading 1-loop diagrams to $\Delta a_\mu$ in the BRpV supersymmetric model
are the same as that in the minimal supersymmetric standard model
(MSSM). In the MSSM the leading contribution to the $\Delta a_\mu$
comes from the neutralino$-$smuon and chargino$-$muon-sneutrino
1-loop diagrams \cite{g-2loop}.
The corresponding 1-loop diagrams in the BRpV supersymmetric model
are such that in the neutralino$-$smuon diagram we have to replace the summation
over neutralinos and smuons by neutralino $\&$ neutrino states and
smuon $\&$ charged Higgs states, respectively. Similar modification
should be done in the diagram of chargino$-$muon-sneutrino to get the analogous diagram
in the BRpV supersymmetric model. However, mixings between the
neutralinos $\&$ neutrinos, charginos $\&$ charged leptons and
sleptons $\&$ Higgs scalars are suppressed due to the smallness of
the neutrino masses.
As a result of this, the additional contribution
due to neutrinos, charged leptons and Higgs bosons is suppressed in
$\Delta a_\mu$ of the BRpV supersymmetric model, and its value
would almost be the same as that in the MSSM.

We neglect the off-diagonal elements in the charged slepton
and sneutrino mass matrices of eq. (\ref{eq:soft}), which
are highly constrained from experimental
data on the flavor changing processes \cite{mueg}. We further
neglect the left-right mixing in the smuons, since it is
proportional to the muon mass.
After neglecting all these various mixing,
below we give the leading analytical formula for $\Delta a_\mu$
in the BRpV supersymmetric model, for which the formula
would be same as that in the MSSM \cite{g-2loop}.
\begin{eqnarray}
\Delta a_\mu &=& \Delta a^{N^0\tilde{\mu}}_\mu + 
\Delta a^{C^\pm\tilde{\nu}_\mu}_\mu,
\nonumber \\
\Delta a^{N^0\tilde{\mu}}_\mu &=& \frac{m_\mu}{16\pi^2}\sum_{A,j}
\left\{\frac{-m_\mu}{6m_{\tilde{\mu}A}^2(1-x_{Aj})^4}
(N^L_{Aj}N^L_{Aj}+N^R_{Aj}N^R_{Aj})
(1-6x_{Aj}+3x_{Aj}^2+2x_{Aj}^3-6x_{Aj}^2\ln x_{Aj}) \right.
\nonumber \\
&& \left.
-\frac{(m_{N^0})_j}{m_{\tilde{\mu}A}^2(1-x_{Aj})^3}N^L_{Aj}N^R_{Aj}
(1-x_{Aj}^2+2x_{Aj}\ln x_{Aj})\right\},
\nonumber \\
\Delta a^{C^\pm\tilde{\nu}_\mu}_\mu &=& \frac{m_\mu}{16\pi^2}\sum_{j}
\left\{\frac{m_\mu}{6m_{\tilde{\nu}_\mu}^2(1-x_j)^4}(C^L_jC^L_j+C^R_jC^R_j)
(2+3x_j-6x_j^2+x_j^3+6x_j\ln x_j) \right.
\nonumber \\
&& \left.
-\frac{(m_{C^\pm})_j}{m_{\tilde{\nu}_\mu}^2(1-x_j)^3}C^L_jC^R_j
(3-4x_j+x_j^2+2\ln x_j)\right\},
\nonumber \\
x_{Aj} &=& \frac{(m_{N^0}^2)_j}{m_{\tilde{\mu}A}^2}, \quad
x_j = \frac{(m_{C^\pm}^2)_j}{m_{\tilde{\nu}_\mu}^2},
\quad
N^L_{Aj} = -y_\mu (U_{0})_{4j}(U_{\tilde{\mu}})_{LA}
-\sqrt{2}g_1(U_{0})_{1j}(U_{\tilde{\mu}})_{RA},
\nonumber \\
N^R_{Aj} &=& -y_\mu (U_{0})_{4j}(U_{\tilde{\mu}})_{RA}
+\frac{1}{\sqrt{2}}(g_2(U_{0})_{2j}+
g_1(U_{0})_{1j})(U_{\tilde{\mu}})_{LA},
\nonumber \\
C^L_j &=& y_\mu (U_{-})_{2j}, \quad
C^R_j = -g_2 (U_{+})_{1j}.
\label{eq:NCcon}
\end{eqnarray}
Here, $m_\mu,y_\mu$ are the mass and Yukawa coupling of muon, respectively.
$U_0$ is a unitary matrix which diagonalizes the neutralino
matrix, given in eq. (\ref{eq:mchim}), as
$(U_0^T M_{\chi^0} U_0)_{ij} = (m_{N^0})_i\delta_{ij}$, where $i,j=1,\cdots,4$.
$U_-$ and $U_+$ diagonalizes the chargino matrix
\begin{equation}
M_C=\left(\begin{array}{cc}
M_2 & g_2v_u \\
g_2v_d & \mu
\end{array}\right)
\end{equation}
as $(U_-^T M_C U_+)_{ij} = (m_{C^\pm})_i\delta_{ij}$, where $i,j=1,2$.
Since we have neglected the left-right mixing in smuon masses,
we take its diagonalizing matrix $U_{\tilde{\mu}}$ to be 2$\times$2 unit matrix.
After neglecting the left-right mixing, the mass
eigenvalues of smuons and muon-sneutrino are
\begin{eqnarray}
m_{\tilde{\mu}1}^2 &=& m_L^2 + m_Z^2\cos 2\beta (\sin^2\theta_W - \frac{1}{2}),
\quad
m_{\tilde{\mu}2}^2 = m_R^2 - m_Z^2\cos 2\beta \sin^2\theta_W,
\nonumber \\
m_{\tilde{\nu}_\mu}^2 &=& m_{\tilde \nu}^2 = m_L^2 + m_Z^2\cos 2\beta \frac{1}{2}.
\end{eqnarray}
Here, $m_L,m_R$ are the soft parameters of left- and right-handed smuons,
which are one of the diagonal mass parameters of the third
and fourth terms of eq. (\ref{eq:soft}).

From eqs. (\ref{eq:tn1l}) and (\ref{eq:a0a1}) it is clear
that the neutrino mass eigenvalues depend on the following
parameters: $M_1,M_2,\mu,\tan\beta,m_L,\epsilon_i,(b_\epsilon)_i,
b_\mu$. From the analytical expression of $\Delta a_\mu$, which is given
in the previous paragraph, we can see that the unknown
parameters on which it depends on are: $M_1,M_2,\mu,\tan\beta,
m_L,m_R$. The common set of parameters on which both the
neutrino masses and the $\Delta a_\mu$ depend on are $M_1,M_2,\mu,
\tan\beta,m_L$. As it is explained in Sec. \ref{s:intro},
since the neutrino masses have been measured accurately and the
experimental value of $a_\mu$ differs from that of the SM
value by about 3$\sigma$, it is interesting to analyze them
in the BRpV supersymmetric model and see what kind of constraints we may
get on the above said common set of parameters. Specifically,
we have scanned over the parameters of $M_1,M_2,\mu,\tan\beta$
and have obtained the allowed area on these parameters, which would
satisfy the neutrino masses and the $\Delta a_\mu$
constraints for definite values of $m_L$, $m_R$ and $b_\mu$. In the scanning
procedure we have given some range of values for the unknown
parameters $\epsilon_i$ and $(b_\epsilon)_i$, which
determine the neutrino mass scale.
In the next section we elaborate on the procedure of our scanning and
present the results.

\section{Analysis and results}
\label{sec:AnR}

To make our analysis simple, we take the mixing angles in
the lepton sector exactly to be the tri-bimaximal pattern as given
in eq. (\ref{eq:tbm}). Then the
unitary matrix which would diagonalize the neutrino mass matrix is
\begin{equation}
U_\nu = \left(\begin{array}{ccc}
\sqrt{\frac{2}{3}} & \frac{1}{\sqrt{3}} & 0 \\
-\frac{1}{\sqrt{6}} & \frac{1}{\sqrt{3}} & \frac{1}{\sqrt{2}} \\
\frac{1}{\sqrt{6}} & -\frac{1}{\sqrt{3}} & \frac{1}{\sqrt{2}}
\end{array}\right).
\end{equation}
Given the mass eigenvalues of the three neutrinos to be $m_1,m_2,m_3$,
and in a basis where the charged lepton mass matrix has been
diagonalized, we get the following 3$\times$3 matrix in the flavor space
\begin{equation}
m_\nu = U^*_\nu \left(\begin{array}{ccc}
m_1 & 0 & 0 \\
0 & m_2 & 0 \\
0 & 0 & m_3 \\
\end{array}\right) U^\dagger_\nu .
\end{equation}
In the BRpV supersymmetric model, we have already obtained
a mass matrix for neutrinos in the flavor space which is given
in eq. (\ref{eq:tn1l}). Equating the above equation to that
in eq. (\ref{eq:tn1l}), we obtain six independent equations.
Solving them consistently we get the following solution
\begin{eqnarray}
&& \epsilon_1=0,\epsilon_2=\epsilon_3=\epsilon,\quad
(b_\epsilon)_1=(b_\epsilon)_2=-(b_\epsilon)_3
=b_\epsilon,
\nonumber \\
&& m_1=0,\quad m_2=3a_1(b_\epsilon)^2,\quad m_3=2a_0\epsilon^2.
\end{eqnarray}
The reason for getting $m_1$ to be zero is that the rank of matrix in
eq. (\ref{eq:tn1l}) is two, which happens due to the assumption
of degenerate sneutrinos.
The above solution suggests that
we can only have hierarchical mass pattern for the neutrinos.
Since we stated before that we consider a
scenario where tree level eigenvalue is the dominant one,
we set the eigenvalue $m_3$ to the atmospheric
scale $\sqrt{|\Delta m_{31}^2|}\approx$ 0.05 eV and the $m_2$
to the solar scale $\sqrt{\Delta m_{21}^2}\approx$ 0.009 eV.
Previously, we have estimated the scales of $\epsilon$ and $b_\epsilon$
to get the right amount of neutrino masses.
In this work, to fit the atmospheric and solar neutrino mass scales, we
have allowed the values of $\epsilon$ to be between $9\times 10^{-4}
-2\times 10^{-3}$ GeV and the range for $b_\epsilon$
to be 100 times the range of $\epsilon$. To put this
statement precisely, the neutrino mass condition with tri-bimaximal
mixing pattern is satisfied if
the supersymmetric point in the parametric space satisfies the
following inequalities
\begin{equation}
9\times 10^{-4}~{\rm GeV} \leq \sqrt{\frac{0.05~{\rm eV}}{2a_0}}
\leq 2\times 10^{-3}~{\rm GeV},\quad
0.09~{\rm GeV}^2 \leq \sqrt{\frac{0.009~{\rm eV}}{3a_1}}
\leq 0.2~{\rm GeV}^2.
\label{eq:nucon}
\end{equation}
The supersymmetric point which satisfies the above
condition should also satisfy the restriction due to the
$\Delta a_\mu$ which is given in eq. (\ref{eq:damu}). We allow a
2$\sigma$ variation in eq. (\ref{eq:damu}), and the
condition for a supersymmetric point to be consistent with
the $(g-2)_\mu$ is to satisfy the following inequality
\begin{equation}
9.1\times 10^{-10} \leq \Delta a_\mu \leq 46.3\times 10^{-10}.
\label{eq:g-2con}
\end{equation}

At the end of previous section we have stated that
our motivation is to scan over the parameters $M_1,M_2,
\mu,\tan\beta$ which would satisfy both the neutrino masses
and the $\Delta a_\mu$ constraints for some fixed values of
$m_L,m_R,b_\mu$ in the BRpV supersymmetric model. The parameters
$b_\mu$ and $\tan\beta$ determine the leading contribution
to the Higgs boson masses and also the angle $\alpha$. In
this work we have used the tree level contributions for the Higgs
related quantities and fixed the parameter $b_\mu = (100~{\rm GeV})^2$.
The scanning on the parameters $M_1,M_2,\mu,\tan\beta$ has been done
over a grid of points in the following different planes:
$M_2-M_1$, $M_2-\mu$ and $M_2-\tan\beta$.
Before presenting our results on the scanning, we mention about
the lower and upper limiting values for the scanned parameters. From the
negative search on supersymmetric particles at the Large Electron
Positron collider, we can set a lower limit of 100 GeV on the
parameters $M_2,\mu$ which determine the chargino masses.
From the naturalness argument
we can put an upper limit on these to be
around 1 TeV. Based on these facts, we take lower and upper
limiting values on the mass parameters $M_1,M_2,\mu$ to be 100 GeV and
1 TeV, respectively. For $\tan\beta$, we scan over positive values
from 1 to 20.

In the scanning on the parameters of neutralinos, we have found
that $\tan\beta$ can be constrained severely from the solar neutrino
mass scale. Hence we first present our scanning results in the plane
$M_2-\tan\beta$. We consider points of the form $(100+m2
\times 20~{\rm GeV},1+t\times 0.2)$ in the plane $M_2-\tan\beta$,
where the integers $m2$ and $t$ vary from 0 to 45 and
0 to 95, respectively. Essentially, the above set of points give
a rectangular grid in the plane $M_2-\tan\beta$, with a step size
of 20 GeV in the $M_2$ and 0.2 in the $\tan\beta$ axes, respectively. At each point
on this plane we vary $M_1$ and $\mu$ from 100 GeV to 1 TeV in steps
of 5 GeV, and verify if the conditions due to neutrino masses, as
given in eq. (\ref{eq:nucon}), and the condition due to $\Delta a_\mu$,
as given in eq. (\ref{eq:g-2con}), are satisfied. We have also done
the above said scanning by checking the constraints due to either
neutrino masses or the $\Delta a_\mu$, to understand the
individual constraints from both these observable quantities.

\begin{figure}[!h]
\begin{center}
\includegraphics[height=2in,width=2in]{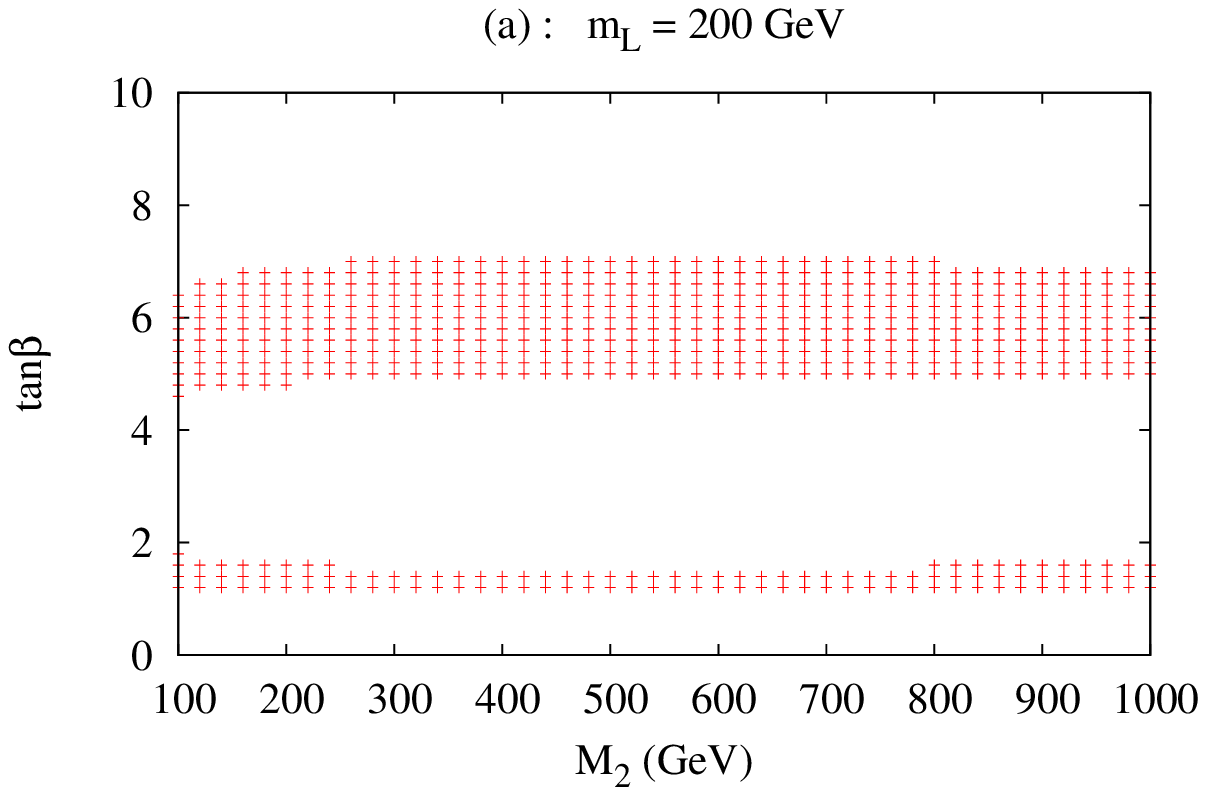}
\includegraphics[height=2in,width=2in]{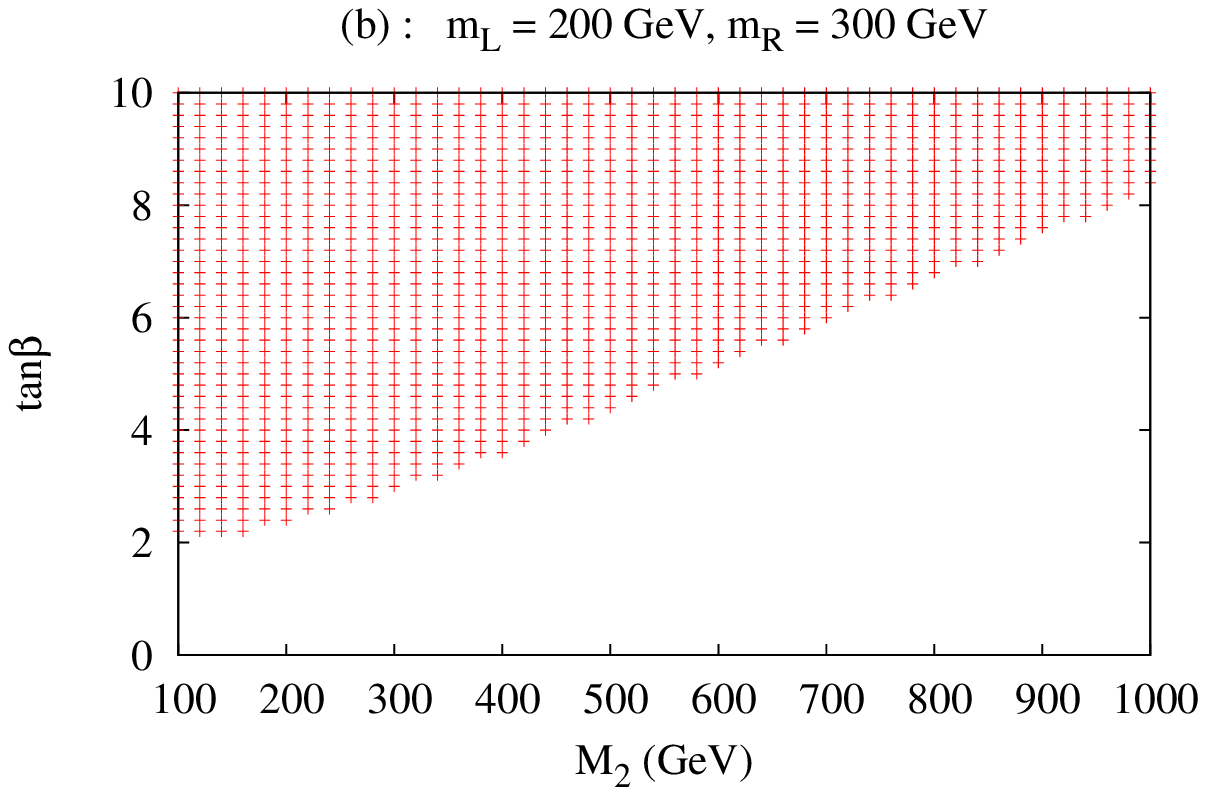}

\includegraphics[height=2in,width=2in]{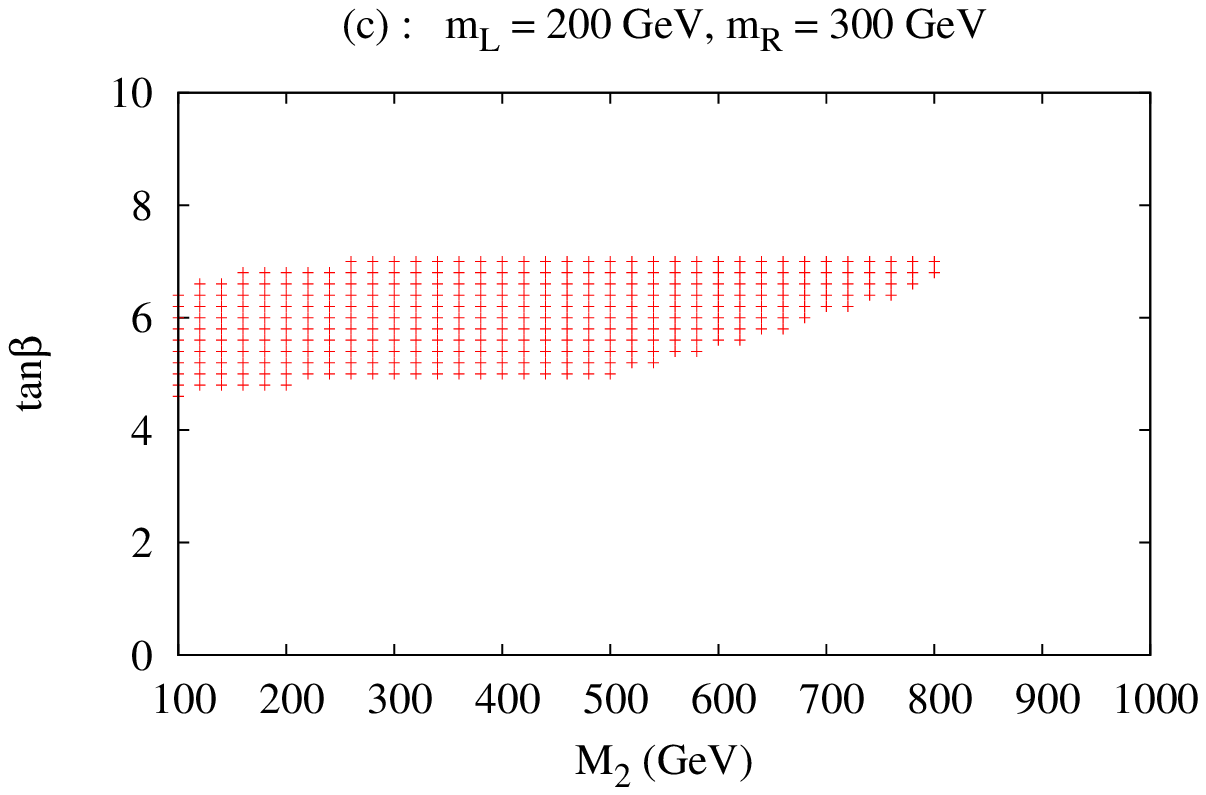}
\includegraphics[height=2in,width=2in]{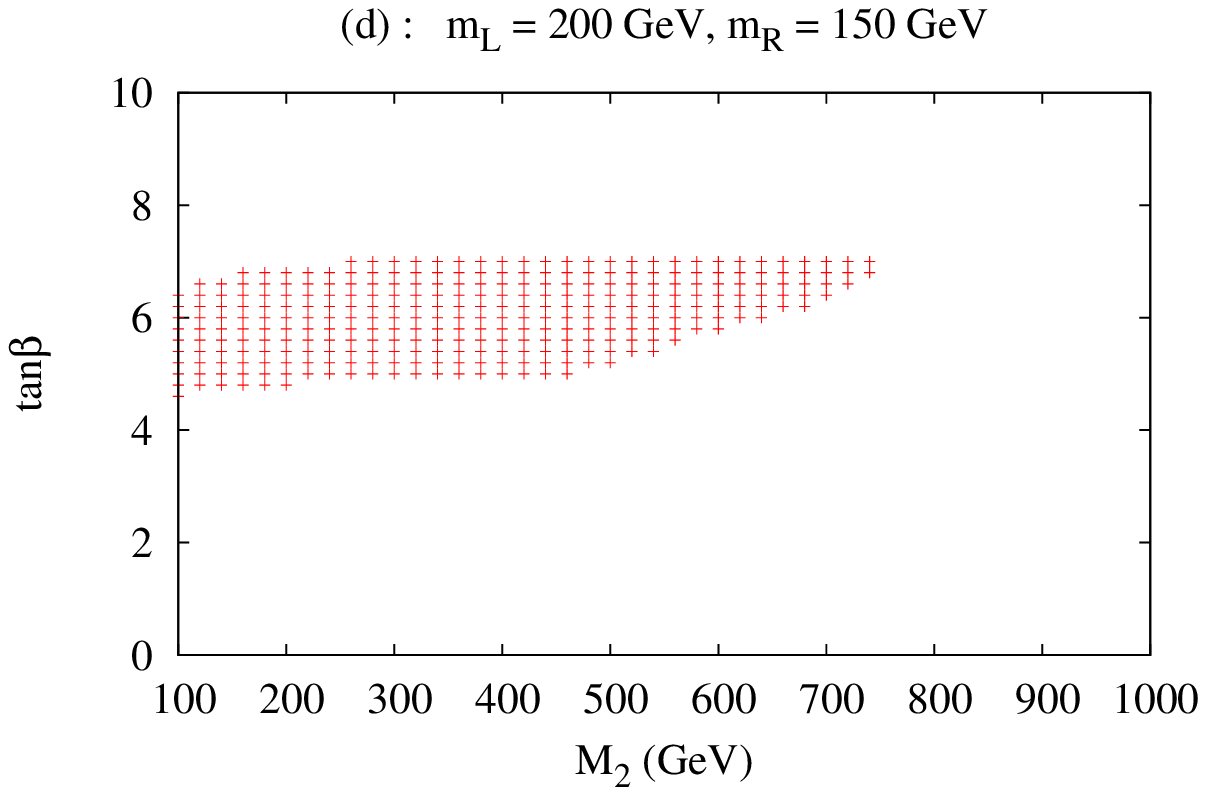}
\end{center}
\vspace*{-1cm}
\caption{Allowed space in the plane $M_2-\tan\beta$. In the upper-left
and upper-right plots, only the neutrino mass
and $\Delta a_\mu$ constraints have been applied, respectively. In
the lower-left and lower-right plots, both the neutrino mass
and $\Delta a_\mu$ constraints have been applied for different set
of soft parameters $m_L,m_R$. For details, see the text.}
\label{f:2m2tb}
\end{figure}

In Fig. \ref{f:2m2tb}(a) we have given allowed set of points
in the plane $M_2-\tan\beta$ for $m_L$ = 200 GeV by the neutrino mass
constraint of eq. (\ref{eq:nucon}). The $\tan\beta$
is restricted to be less than about 7, and also, no points are allowed in the
plane for $\tan\beta$ between about 2 and 4. These stringent limits on the $\tan\beta$
are coming from the solar neutrino mass scale rather than from the
atmospheric scale. Numerically, we have seen that for any fixed
values of $M_1,M_2,\mu$, the quantity $a_1$, which is defined in eq. (\ref{eq:a0a1}),
increases with $\tan\beta$ in magnitude upto $\tan\beta\sim 3$
and then decreases afterwards. We have found that at $\tan\beta\sim 3$,
the sneutrino mass is almost degenerate with the heavy Higgs and
the pseudo-scalar Higgs masses, and hence, the
quantity $a_1$ would be enhanced due to singularity in the
function $I_4$ of eq. (\ref{eq:i4}). For $\tan\beta$ between about
2 to 4 the magnitude of $a_1$ is so high that the second inequality
of eq. (\ref{eq:nucon}) is not satisfied. After reaching the maximum
at $\tan\beta\sim$ 3, the quantity $a_1$ would decrease in magnitude
with increasing $\tan\beta $, due to increase in the masses of heavy
and pseudo-scalar Higgses. So, for $\tan\beta$ greater than 7, the
magnitude of $a_1$ would come out to be so less that the second
inequality in eq. (\ref{eq:nucon}) is not satisfied.

For $m_L$ = 200 GeV and $m_R$ = 300 GeV, we have given the allowed
points in the plane $M_2-\tan\beta$ due to the $\Delta a_\mu$ constraint of
eq. (\ref{eq:g-2con}) in Fig. \ref{f:2m2tb}(b).
From the figure it can be noticed that by increasing the
$M_2$ values the lower limit on $\tan\beta$ is increasing. This fact can be
understood from the dependences of $\Delta a_\mu$ on relevant parameters.
Numerically,
we have seen that the $\Delta a_\mu$ is inversely proportional to either
$M_2$ or $\mu$ and it is directly proportional to $\tan\beta$ \cite{tb}.
For $m_R > m_L$, which is the case in Fig. \ref{f:2m2tb}(b),
the dependence of $M_1$ on $\Delta a_\mu$ is negligible, since the
$\Delta a_\mu^{C^\pm\tilde{\nu}_\mu}$ of eq. (\ref{eq:NCcon})
would dominate over the $\Delta a_\mu^{N^0\tilde{\mu}}$.
Now, it is easy to
understand that by increasing $M_2$ the value of $\Delta a_\mu$ may
come below the lower limit of eq. (\ref{eq:g-2con}), which
can be compensated by increasing $\tan\beta$, and thus
we get the allowed space as shown in Fig. \ref{f:2m2tb}(b).

In Fig. \ref{f:2m2tb}(c) we have given allowed points by both the
eqs. (\ref{eq:nucon}) and (\ref{eq:g-2con}) in the plane $M_2-\tan\beta$
for $m_L$ = 200 GeV and $m_R$ = 300 GeV. Essentially, the points in
Fig. \ref{f:2m2tb}(c) are almost
the intersection of the points of Figs. \ref{f:2m2tb}(a) and
\ref{f:2m2tb}(b). By combining the neutrino mass constraints
with the constraint due to the $\Delta a_\mu$, we can get upper bound on
the $M_2$ which is shown in Fig. \ref{f:2m2tb}(c).
In Fig. \ref{f:2m2tb}(d) we have given allowed points by both the eqs.
(\ref{eq:nucon}) and (\ref{eq:g-2con}) in the plane $M_2-\tan\beta$
for $m_L$ = 200 GeV and $m_R$ = 150 GeV. The value of $m_R$ in Fig.
\ref{f:2m2tb}(d) is less than that in Fig. \ref{f:2m2tb}(c).
The parameter $m_R$ affect the $\Delta a_\mu^{N^0\tilde{\mu}}$
of eq. (\ref{eq:NCcon}). Numerically,
we have seen that the quantity $\Delta a_\mu^{N^0\tilde{\mu}}$ would give
negative contribution
and its magnitude is less than that of $\Delta a_\mu^{C^\pm\tilde{\nu}_\mu}$
which gives positive values. By decreasing the value of $m_R$
the negative contribution due to neutralinos would increase in
magnitude, and as a result the net $\Delta a_\mu$ decreases from
its previous value. Since the $\Delta a_\mu$ also decreases with increasing
the $M_2$, we loose 
some more points on the higher end of $M_2$, which can be seen by comparing
Fig. \ref{f:2m2tb}(d) with Fig. \ref{f:2m2tb}(c).

\begin{figure}[!h]
\begin{center}
\includegraphics[height=2in,width=2in]{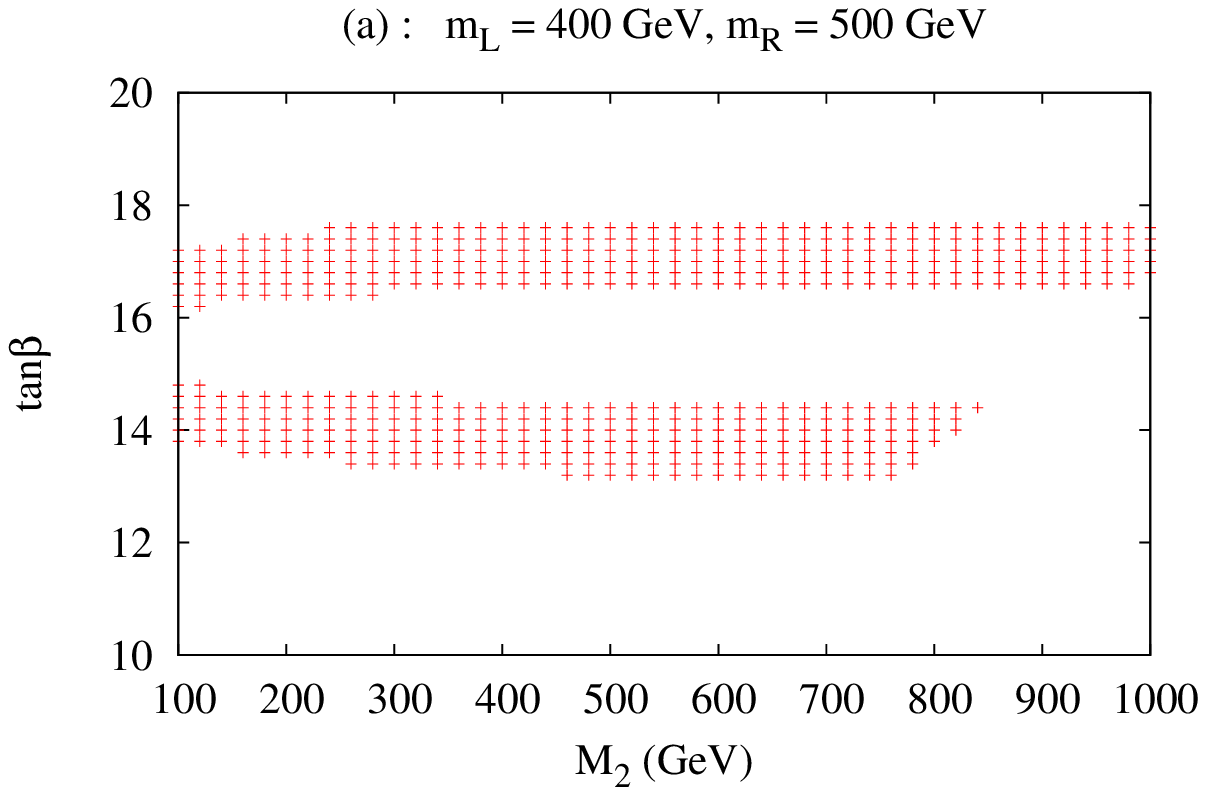}
\includegraphics[height=2in,width=2in]{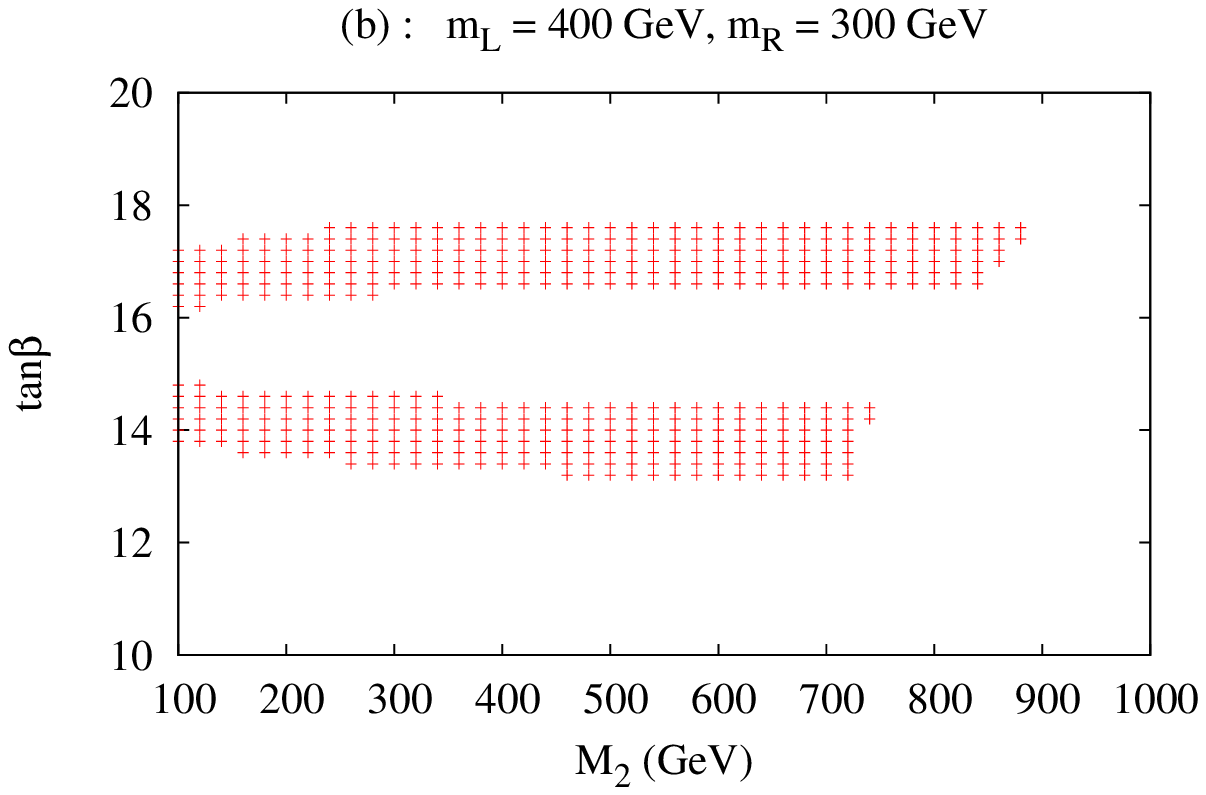}
\end{center}
\vspace*{-1cm}
\caption{Allowed space in the plane $M_2-\tan\beta$ by both
the neutrino mass and the $\Delta a_\mu$ constraints. The
values of $m_L,m_R$ are
higher in this case as compared to that in Fig. \ref{f:2m2tb}.}
\label{f:4m2tb}
\end{figure}

In the plots of Fig. \ref{f:2m2tb}, the values of $m_L$ and
$m_R$ are somewhat low. By increasing these values we have
done a similar scanning in the plane $M_2-\tan\beta$ to
illustrate how the constraints would change quantitatively.
In Figs.
\ref{f:4m2tb}(a) and \ref{f:4m2tb}(b) we have applied
constraints due to eqs. (\ref{eq:nucon}) and (\ref{eq:g-2con}),
for $(m_L,m_R)$ = (400 GeV, 500 GeV) and (400 GeV, 300 GeV),
respectively. The allowed $\tan\beta$ in these plots is between
about 13 and 18, whose values are high compared to that in
Fig. \ref{f:2m2tb}. For $\tan\beta$ either larger than 18 or
lesser than 13 the quantity $a_1$ is suppressed,
and for $\tan\beta$ between about 15 and 16 the $a_1$ is enhanced,
so that the second inequality of eq. (\ref{eq:nucon}) is
not satisfied. The reasons for suppression or enhancement of
$a_1$ in this case is analogous to the case of Fig. \ref{f:2m2tb}(a).
On top of these
neutrino mass constraints, the constraint due to $\Delta a_\mu$
can eliminate points on the higher end of $M_2$, which is
evident in Fig. \ref{f:4m2tb}(a).
The effect of negative enhanced contribution of the neutralino
diagram to $\Delta a_\mu$, which we explained in the
previous paragraph, can be seen in Fig. \ref{f:4m2tb}(b).

We now give the results of scanning in the plane $M_2-\mu$.
In order to have a grid of points
in this plane, we consider points of the form $(100+m2\times
20~{\rm GeV},100+m\times 20~{\rm GeV})$, where the integers
$m2,m$ vary independent of one another from 0 to 45. Basically,
these set of points form a grid in the plane $M_2-\mu$
with a step size of 20 GeV in both the $M_2$ and $\mu$ axes.
Since we have known the allowed $\tan\beta$ from Figs. \ref{f:2m2tb}
and \ref{f:4m2tb}, at each
point on the grid of the plane $M_2-\mu$,
we run $M_1$ from 100 GeV to 1 TeV in steps of 5 GeV and vary $\tan\beta$
either from 1 to 7 in steps of 0.2 if $m_L$ is 200 GeV or from 13 to 18
in steps of 0.2 if $m_L$ is 400 GeV.

\begin{figure}[!h]
\begin{center}
\includegraphics[height=2in,width=2in]{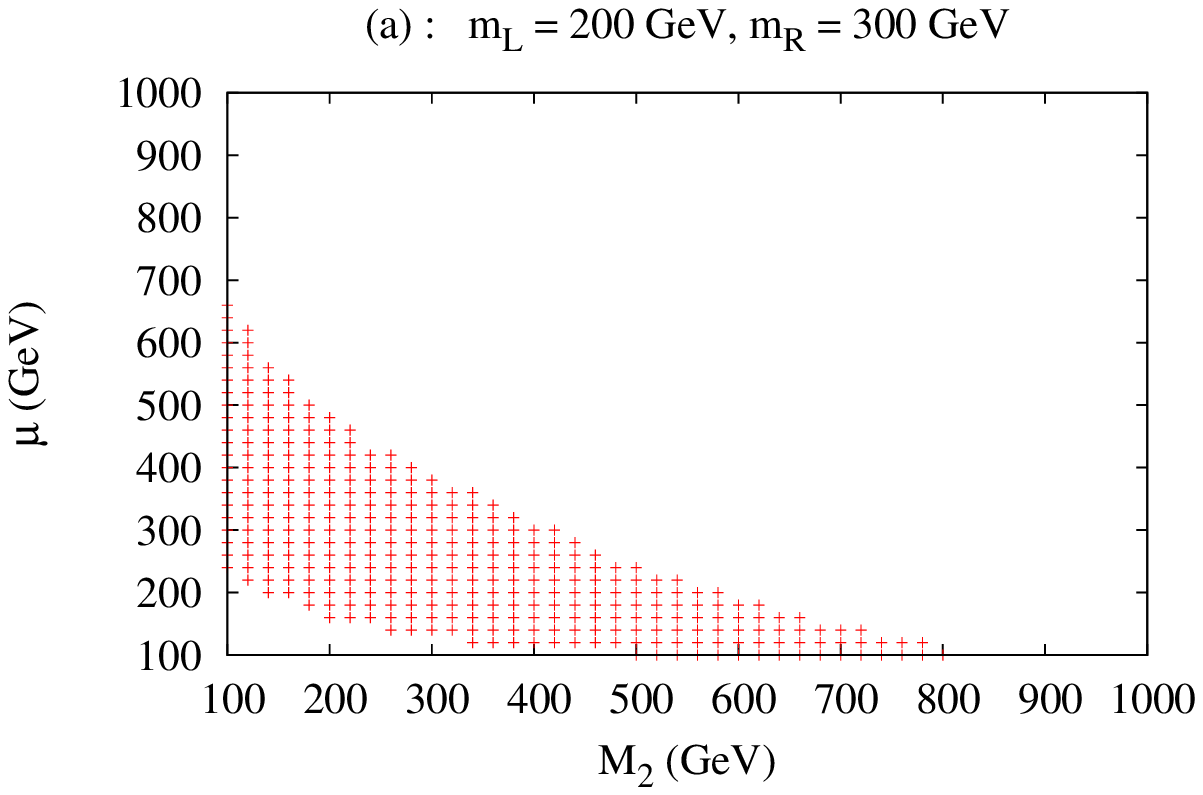}
\includegraphics[height=2in,width=2in]{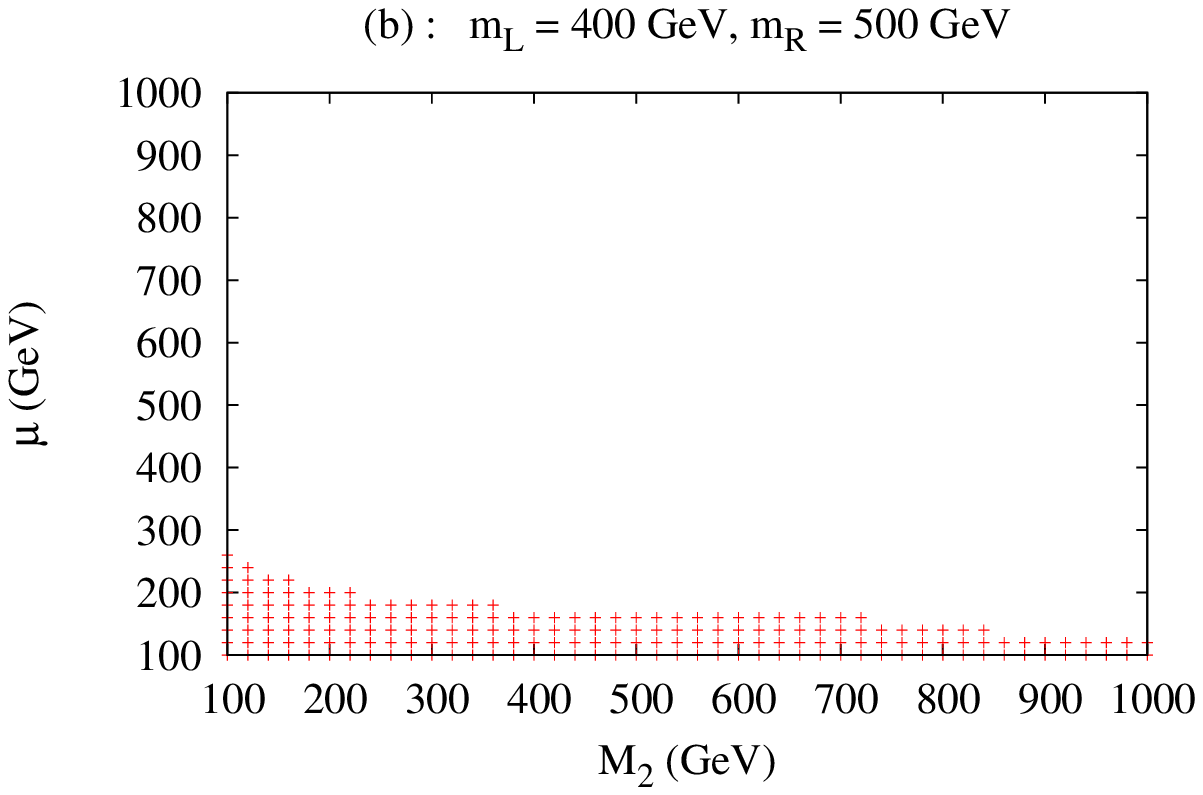}
\end{center}
\vspace*{-1cm}
\caption{Allowed space in the plane $M_2-\mu$ by both the
neutrino mass and the $\Delta a_\mu$ constraints for two different
set of $(m_L,m_R)$ values.}
\label{f:2m2mu}
\end{figure}

In Fig. \ref{f:2m2mu}(a) we have given allowed points by both
the neutrino mass and the $\Delta a_\mu$ constraints for
$(m_L,m_R)=$ (200 GeV, 300 GeV).
In this plot, the white area in the
lower-left corner where both $M_2$ and $\mu$ are
small is disallowed because the tree level neutrino
mass eigenvalue cannot fit the atmospheric scale.
The atmospheric scale is determined by the quantity
$a_0$ which decreases with increasing either $\mu$ or
$\tan\beta$, which can be seen from eq. (\ref{eq:a0a1}).
Since the
solar scale demands that the $\tan\beta$ to be less than about 7 for
$m_L$ = 200 GeV, the quantity $a_0$ would be so high for low values of
$\mu,M_2$. These high values of $a_0$ do not
satisfy the first inequality of eq. (\ref{eq:nucon}). The vast white
area on the top-right side of the allowed narrow strip is disallowed
because the values of $\Delta a_\mu$
are coming out to be smaller than the lower limit of eq. (\ref{eq:g-2con}),
which can be understood from the dependences of $\Delta a_\mu$
on the relevant parameters.
%

In Fig. \ref{f:2m2mu}(a) $m_L$ is fixed to 200 GeV. The
results of scanning over the plane $M_2-\mu$ can dramatically
change if we increase $m_L$ to 400 GeV. In Fig. \ref{f:2m2mu}(b)
we have given the allowed points by both the neutrino mass and
the $\Delta a_\mu$ constraints for $(m_L,m_R)$ = (400 GeV, 500 GeV).
In this plot,
the $\mu$ is constrained to be within 260 GeV. The allowed $\tan\beta$
range in this case is somewhat large and
gives suppression in the quantity $a_0$ of eq. (\ref{eq:a0a1}).
The $a_0$ further gets suppression if $\mu$ is larger than 260 GeV
so that the first inequality of eq. (\ref{eq:nucon}) is not satisfied.
Unlike in Fig. \ref{f:2m2mu}(a),
the role of $\Delta a_\mu$ constraint is very minimal
and the majority of constraints in Fig. \ref{f:2m2mu}(b)
are due to neutrino masses. In both the plots
of Fig. \ref{f:2m2mu} the parameter $m_L$ is less than that of $m_R$.
By setting $m_L>m_R$, we may see a reduction in the number of
points in the higher end of $M_2$, due to
negative enhanced neutralino contribution of $\Delta a_\mu$.

\begin{figure}[!h]
\begin{center}
\includegraphics[height=2in,width=2in]{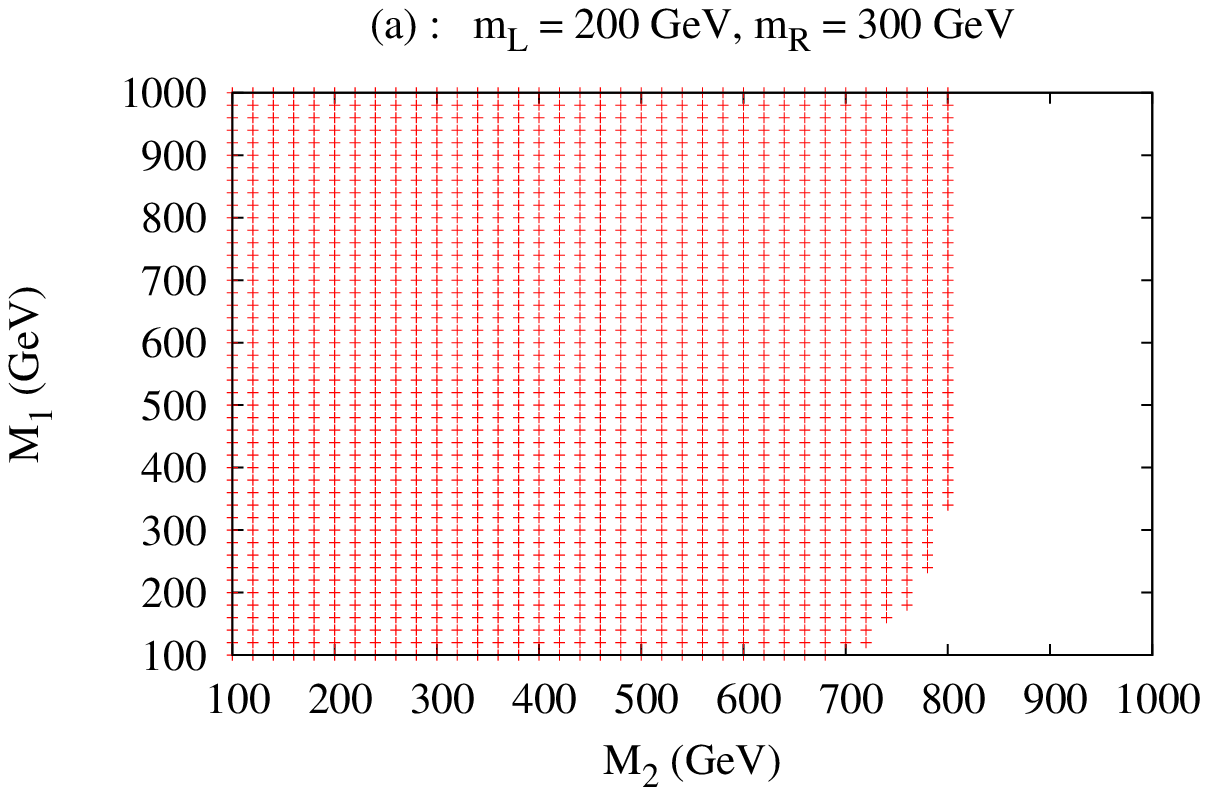}
\includegraphics[height=2in,width=2in]{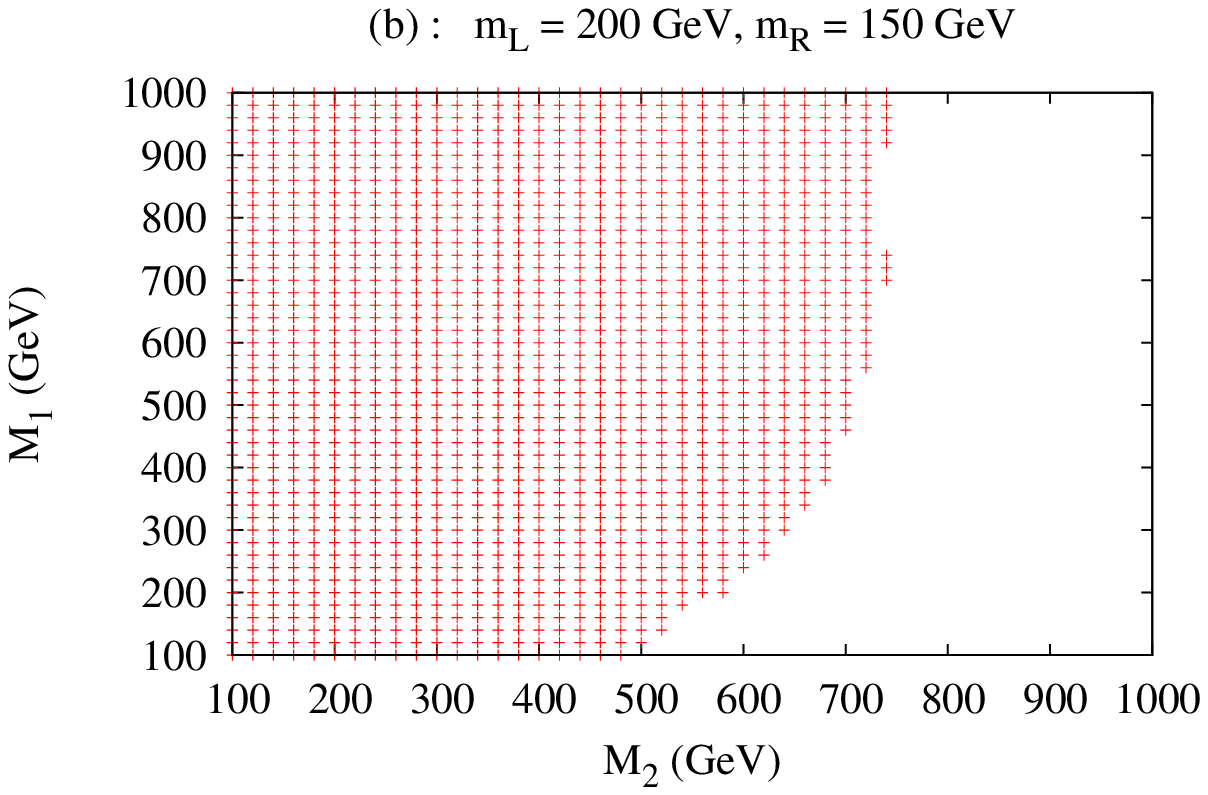}
\end{center}
\vspace*{-1cm}
\caption{Allowed space in the plane $M_2-M_1$ by both the neutrino
mass and the $\Delta a_\mu$ constraints for two different sets
of $(m_L,m_R)$ values.}
\label{f:2m2m1}
\end{figure}

Our method of doing scanning in the plane $M_2-M_1$ is similar to
what we have done in the plane $M_2-\mu$. If we replace $\mu$ with
$M_1$ and vice versa in the scanning of $M_2-\mu$, which we have
described previously, this would give scanning in the plane $M_2-M_1$.
In Figs. \ref{f:2m2m1}(a) $\&$ \ref{f:2m2m1}(b) we have given
allowed parametric space in this plane due to neutrino masses
and the $\Delta a_\mu$ constraints for $(m_L,m_R)$ = (200 GeV, 300 GeV)
and (200 GeV, 150 GeV), respectively. The exclusion
area in these plots, which is shown in white area, can be understood
from the properties of the neutrino masses and the $\Delta a_\mu$, which
we have described previously. Although there is some
exclusion area in Figs. \ref{f:2m2m1}(a) $\&$ \ref{f:2m2m1}(b),
there is no upper bound on the $M_1$ parameter unlike an upper bound on the
$M_2$, which can be seen in the above plots. This implies that the
parameter $M_1$ is not sensitive to either of the neutrino
mass eigenvalues and to the $\Delta a_\mu$ as well. We have repeated
the above scanning in the plane $M_2-M_1$ by increasing the $(m_L,m_R)$
values. Even in this case we have found that there could be some exclusion
area in the above plane but no stringent bounds on the parameter
$M_1$.

%

One comment on our scanning results is that we have fixed
the soft parameter of the bilinear Higgs term $b_\mu$ to be $(100~{\rm GeV})^2$,
which determines the Higgs boson masses at tree level.
By changing this parametric value,
the white disallowed area in Figs. \ref{f:2m2tb} and \ref{f:4m2tb} which
happens due to the degeneracy of the sneutrino mass and the Higgs boson
masses, may change quantitatively. For a related work on the
consistency of neutrino masses and $(g-2)_\mu$, see \cite{FKO}.

\begin{figure}[!h]
\begin{center}
\includegraphics[height=2in,width=2in]{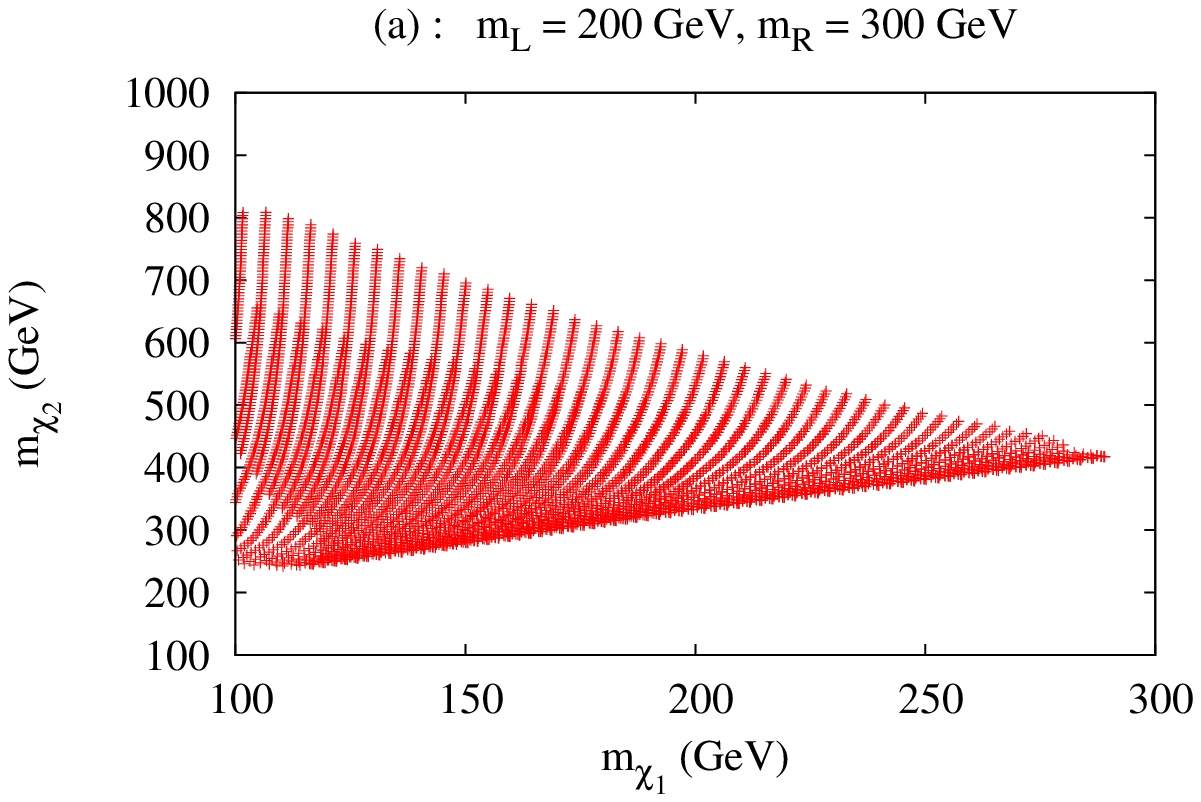}
\includegraphics[height=2in,width=2in]{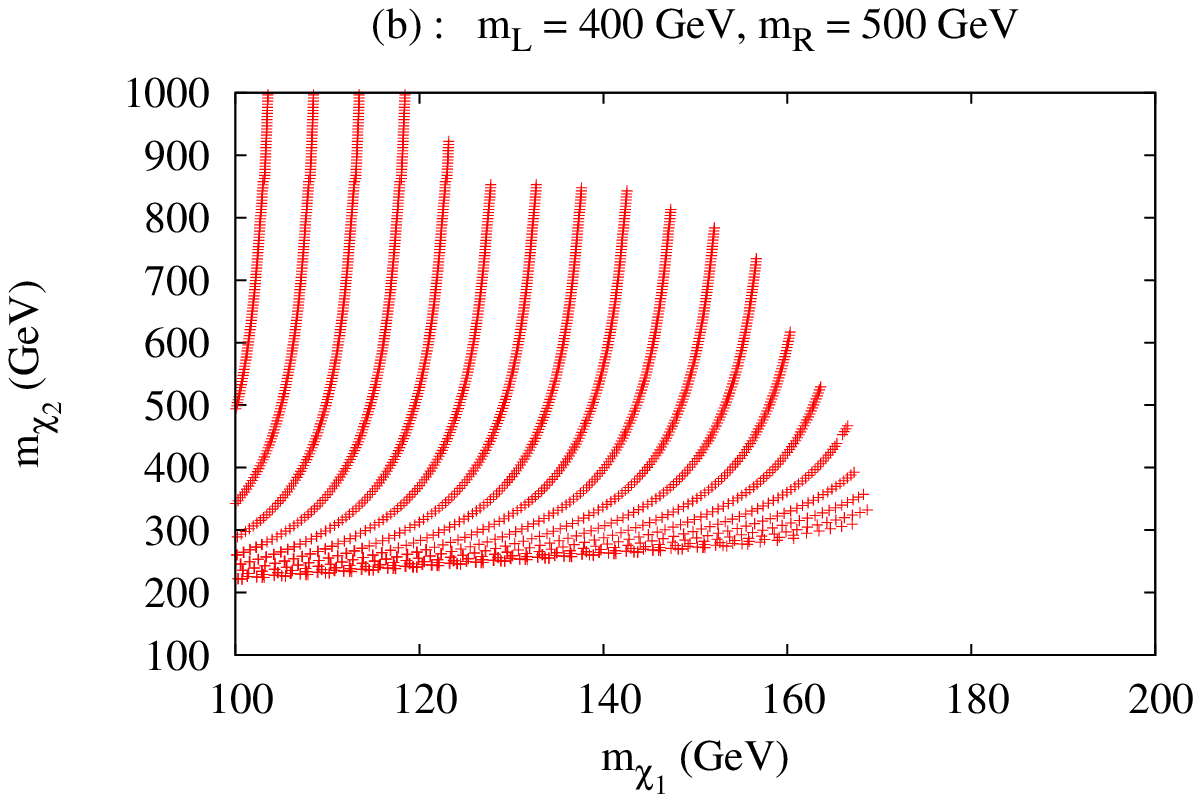}
\end{center}
\vspace*{-1cm}
\caption{Lightest versus heaviest chargino mass plot, which is allowed
by the constraints due to neutrino masses and the $\Delta a_\mu$. $m_{{\chi}_1}$
and $m_{{\chi}_2}$ are the lightest and the heaviest chargino
masses, respectively. For details, see the text.}
\label{f:cha}
\end{figure}

Finally, we give one of the applications of our scanning procedure. One
of the main channels to probe supersymmetry in experiments
is the production of chargino and its decay to lighter particles. In this
context it would be interesting if we can find some upper bounds on the
chargino masses in the present model. The tree level chargino masses are
determined by the following parameters: $M_2$, $\mu$ and $\tan\beta$.
In Figs. \ref{f:2m2mu}(a) $\&$ \ref{f:2m2mu}(b), we have given the allowed
parametric space in the plane $M_2-\mu$ due to both the neutrino masses
and the $\Delta a_\mu$ constraints.
Since we know the $\tan\beta$ range in both these
plots, we can calculate the lowest and the heaviest chargino
masses at each allowed point of the plane $M_2-\mu$.
To get the results for
chargino masses we take the grid step size to be 5 GeV in the $M_2$ and $\mu$ axes.
At each point on this grid we vary $M_1$ and $\tan\beta$ as it has been
done in Figs. \ref{f:2m2mu}(a) $\&$ \ref{f:2m2mu}(b). We also fix the necessary soft
parameters to the values as they are taken in Figs. \ref{f:2m2mu}(a) $\&$ \ref{f:2m2mu}(b).
In Fig. \ref{f:cha} we
have given the plots for the lightest ($m_{\chi_1}$) versus heaviest ($m_{\chi_2}$) chargino masses
which are allowed by the neutrino masses and the $\Delta a_\mu$ constraints.
The empty space between some contour kind of lines in these plots
is happening due to finite grid step size of the plane $M_2-\mu$.
From these plots we can see that for a definite value of lightest
chargino mass we can get an upper and lower bounds on the
heaviest chargino mass, and vice versa. In Fig. \ref{f:cha}(a)
for $m_{\chi_1}$ = 150 GeV, the lower and upper bounds on the heaviest
chargino mass are 270 GeV and 700 GeV, respectively. Whereas, in Fig.
\ref{f:cha}(b) for the same $m_{\chi_1}$ = 150 GeV, the value of $m_{\chi_2}$
lies between about 250 GeV and 800 GeV. Also, depending on
the values of the soft masses we can get an absolute upper bound on
the lightest chargino mass. In Fig. \ref{f:cha}(a), the absolute
upper bound on the lightest chargino is about 290 GeV, whereas,
in Fig. \ref{f:cha}(b) it is about 170 GeV.

\section{Conclusions}
\label{sec:con}

The smallness of neutrino masses and the 3$\sigma$ deviation of the experimental
value of the muon $(g-2)_\mu$ from that of the standard model value, may indicate
new physics. We have studied how the observed values of these two quantities
put constraints on some parametric values in the bilinear R-parity violating
supersymmetric model. One interesting feature
of this model is that the parameters which determine the neutralino and chargino
masses also determine the neutrino masses as well as the $\Delta a_\mu$. The
neutrino masses and the $\Delta a_\mu$ depend on some soft parameters as well.
By taking
these soft parameters to some fixed values, we have scanned over the following
parameters: $M_1$, $M_2$, $\mu$ and $\tan\beta$, and presented allowed
parametric space by satisfying the observed data on the neutrino masses
and the $\Delta a_\mu$.
For neutrino mixing angles, we have assumed the tri-bimaximal
mixing pattern.

The plots we have presented on the allowed parametric space of the
parameters, $M_1$, $M_2$, $\mu$ and $\tan\beta$, are for some
specific values of the soft parameters and also for some allowed range
of the bilinear parameters $\epsilon$ and $b_\epsilon$. In this sense
the plots we have given here are only indicative. But, we have
described
some generic features of the neutrino masses and the $\Delta a_\mu$ and
how they play a part in setting bounds on the model parameters.
Specifically, we have found that in the BRpV supersymmetric
model, depending on the soft parametric values, the constraints
due to neutrino masses and the $\Delta a_\mu$ can put certain
limits on the parameters $M_2,\mu,\tan\beta$,
but not on the parameter $M_1$.
One of the uses of our scanning procedure
is that the allowed parametric space we get on the
$M_2$, $\mu$ and $\tan\beta$,
can set upper and lower bounds on the chargino masses
of this model.
Indeed, given the data from collider experiments, the results of
this scanning procedure can be used to probe the
supersymmetric particle spectrum of this model.

\section*{Acknowledgments} The author wishes to thank Xerxes Tata
for having some discussions in the initial stages of this work.
The author is grateful to Sourov Roy for valuable discussions and suggestions
on this work and also for reading the manuscript.

\end{document}